\documentstyle[aps,prl,epsf,multicol]{revtex}
\begin{document} 
\draft 
\title{Coulomb blockade in metallic grains at large conductance: beyond
  the zero-dimensional limit} 
\author{I.S. Beloborodov and A.V. Andreev} 
\address {Bell Laboratories, Lucent Technologies, Murray Hill, NJ 07974 \\
Department of Physics, University of Colorado, CB 390, Boulder, CO
80390}

\date{\today} \maketitle

\begin{abstract}
  We consider mesoscopic fluctuations of the Coulomb blockade
  oscillations in a disordered metallic grain in an external magnetic
  field coupled to a metallic lead by a tunneling contact with a large
  conductance $g_T$ and capacitively coupled to a gate.  For a finite mean
  level spacing $\delta$ in the grain and at temperatures $T \gg \delta$
  we obtain the 
  oscillatory part of the correlator of thermodynamic potentials at 
  different values of the gate  voltage. The correlations
  decay algebraically with the  gate voltage difference in contrast to
  the exponential decay that would result from the mean-field treatment
  of the Coulomb interaction.  The results are valid beyond the 
  zero-dimensional limit and describe the crossover between the unitary
  and orthogonal ensembles.  
\end{abstract}


\begin{multicols}{2}

\begin{figure}
  \epsfysize =4cm \centerline{\epsfbox{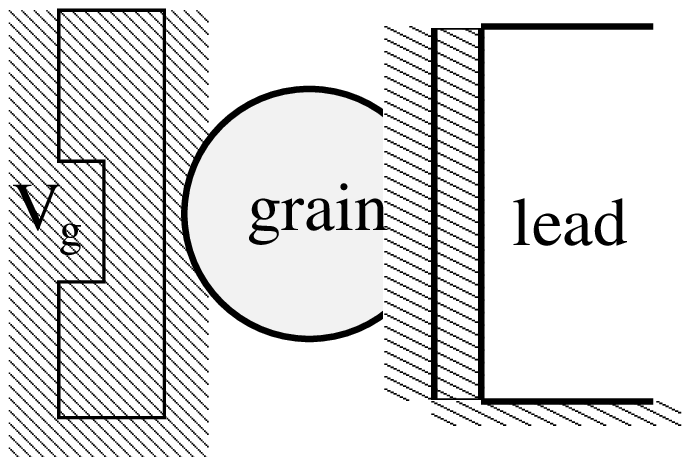}}
\caption{Schematic drawing of a disordered metallic grain coupled by a
  tunneling contact to a metallic lead. The charge of the grain is
  controlled by the gate voltage $V_g\sim q$}
\label{dot}
\end{figure}

  Coulomb blockade is one of the most striking manifestations of
  electron interactions at low temperatures~\cite{Kouwenhoven}. It can
  be observed by measuring, say the charge of a small metal particle
  connected by a tunneling contact to a metallic lead and capacitively
  coupled to a metallic gate which is maintained at voltage $V_g$, as
  in Fig.~\ref{dot}.
  
  In the weak tunneling regime the charge in the grain is nearly
  quantized at low temperatures and exhibits a characteristic
  step-like dependence on the gate voltage $V_g$. This leads to the
  Coulomb blockade  oscillations in the gate voltage dependence of
  various physical quantities.  The amplitude of these oscillations
  decreases with increasing transparency of the contact and depends
  not only on the total conductance of the contact but on the
  individual transparencies of the tunneling channels.
  
  In the strong tunneling regime most of the studies of the Coulomb
  blockade  concentrated on the limit of large grains with the vanishing
  mean level spacing $\delta$. In this limit the Coulomb blockade
  oscillations of the thermodynamic quantities vanish at perfect
  transmission~\cite{Flensberg93,Matveev95}.  In the case of a
  multi-channel tunneling contact  the Coulomb blockade oscillations at
  large conductance  have been shown to be exponentially small in the
  dimensionless tunneling conductance $g_T$~\cite{Zaikin,Grabert}.  The 
  case of a diffusive random contact  was considered in
  Refs.~\cite{Nazarov,Kamenev}.  At $\delta=0$ the mesoscopic fluctuations
  of the Coulomb blockade oscillations due to disorder in the grain vanish.

  The case of strong tunneling and finite $\delta$
  was studied in Ref.~\cite{Aleiner97} where it was shown that the Coulomb
  blockade oscillations  remain finite  even at perfect transmission and
  exhibit strong mesoscopic  fluctuations due to the presence of disorder
  in the grain.  This study utilized the bosonization  method of
  Ref.~\cite{Matveev95}  which  is applicable to contacts with weakly
  reflecting channels.

  Recently, much interest was focussed on granulated disordered
  metals~\cite{Efetov,experiment}, which can be viewed as extended
  networks of metallic grains in the strong  tunneling regime.  The low
  temperature regime, where mesoscopic fluctuations at finite $\delta$
  become observable is
  well within experimental reach~\cite{experiment}. It is therefore
  desirable to generalize the study of charging effects at finite
  $\delta$ and strong tunneling~\cite{Aleiner97} to granulated metals.   
  The formalism of Refs.~\cite{Matveev95,Aleiner97} is well suited to
  treat individual grains but is difficult to generalize to extended
  disordered systems.  On the other hand such systems are conveniently
  described within the $\sigma$-model formalism.  It is therefore important
  to incorporate the non-perturbative Coulomb blockade effects into the
  framework of the $\sigma$-model approach.  

In this Letter we show that mesoscopic fluctuations of the Coulomb
blockade oscillations in a metallic grain with finite $\delta$ in the
strong tunneling regime can be studied using the $\sigma$-model. 
We concentrate on the thermodynamic quantities of a  grain of size $L$ with
  the Thouless energy $E_T=D/L^2$, where $D$ is the diffusion constant,
placed in an external magnetic field $H$. 
  The grain is connected to a metallic lead by a non-random multi-channel
  tunneling contact with the dimensionless conductance $g_T = 2\pi \hbar
  /e^2 R \gg 1$ where $R$ is the resistance of the contact. It is also 
  capacitively coupled to a gate, as in Fig.~\ref{dot}.  We assume
  that the electron-electron interaction in the lead may be neglected
  due to screening and treat the Coulomb interaction of electrons in
  the grain within the framework of the constant interaction model,
\begin{equation}
  \label{eq:cint} 
  \hat{H}_C=E_C \left(\hat{N}-q\right)^2. 
\end{equation}
Here $E_C$ is the charging energy, $\hat{N}$ is the operator of the number
of electrons in the grain, and $q \propto V_g $ is the number of
electrons that minimizes the electrostatic energy of the grain.
We consider the case $E_T < E_C$, and describe the crossover
between the unitary and orthogonal ensembles. The case $E_T >
E_C$ for the unitary ensemble was recently studied in
Ref.~\cite{Beloborodov} by a different method.

The single particle levels in the grain are broadened due to tunneling
into the lead.  We assume that tunneling contact is broad, so that
each state $x$ in the dot is coupled to many lead states $a$.  In this
case the probability distribution $P(\Gamma_x)$ of level half-widths
$\Gamma_x$ is sharply peaked about the mean value $\Gamma$.  We can
therefore neglect the fluctuations of $P(\Gamma_x)$ and consider them
to be equal to the mean value $\Gamma$ which can be expressed through
the dimensionless conductance of the contact $g_T$ as
$\Gamma=\case{g_T\delta}{8\pi}$.  The temperature is assumed to
satisfy the conditions $\delta \ll T \ll E_C$.

The main results of this paper are the following expressions for the
oscillatory parts of the {\em ensemble averaged} thermodynamic 
potential $\langle \Omega(q)\rangle_{\rm osc}$ and of the 
irreducible  correlator
$\langle\langle\Omega(q)\Omega(q')\rangle\rangle_{\rm osc}=
\langle\Omega(q)\Omega(q')\rangle_{\rm osc}-
\langle\Omega(q)\rangle_{\rm osc}\langle\Omega(q')\rangle_{\rm osc}$ at
$\delta |q-q'| \gg \Gamma + T$, 
\begin{mathletters}
\label{2}
    \begin{eqnarray}
      \label{eq:onepoint}
 \langle\Omega(q)\rangle_{\rm osc}= - \tilde{E}_C
\ln\left[\frac{E_C}{T+\Gamma}\right] \cos(2\pi q), \\[1ex] 
\label{irreducible}
\langle\langle\Omega(q)\Omega(q')\rangle\rangle_{\rm osc} = \frac{1}{2}
\tilde{E}_C^2 \Re \sum\limits_{\rho=0,1} \left[ e^{2\pi i(q-q') 
}B_\rho \right].
    \end{eqnarray}
Here $\tilde{E}_C=\case{g_T^2E_C}{\pi^2}e^{-g_T/2}$ and $B_\rho$ can be
conveniently expressed through the quantity 
$a_{\rho}=i\delta( q-q')/2+\rho/\tau_H$, with the cooperon gap $1/\tau_H
=DeH/\hbar c$,
\begin{equation}
\label{det3}
B_\rho =\left\{
\begin{array}{cl}
\left(\frac{4\Gamma}{a_{\rho}}\right)^2\ln^2\frac{a_{\rho}}{\Gamma}, 
&   | a_\rho |\ll E_T ,     \\    
 \frac{4^{3-d}\Gamma^2}{\pi E_T^{d/2}a_{\rho}^{2-d/2}}
\ln^2\frac{a_{\rho}}{\Gamma}
, &  |a_\rho| \gg E_T. 
\end{array}
\right.
\end{equation}
  \end{mathletters}
Here  the index $\rho$ labels the diffuson, $\rho=0$, and the
cooperon, $\rho=1$, degrees of freedom and $d=2,3$ is the
dimensionality of the grain.

The results (\ref{2}) were obtained with logarithmic accuracy in the
one instanton approximation and are valid for $T \gg g_T^2
E_C\ln(E_C/\Gamma) \exp(-g_T/2)$.  The Zeeman splitting of energy levels
was neglected. 
Equation (\ref{eq:onepoint}) coinsides with the result for the
zero-dimensional grain~\cite{Beloborodov}. 
At $ q -q'\to \infty$ the function $B_\rho$ in
Eq.~(\ref{det3}) tends to zero, and the correlator $\langle \Omega(q)
\Omega(q')\rangle_{\rm osc}$ factorizes; $\langle \Omega(q)
\Omega(q')\rangle_{\rm osc} = \langle \Omega(q)\rangle_{\rm osc}
\langle \Omega(q')\rangle_{\rm osc} $.

Below we outline the derivation of Eq.~(\ref{2}).  We consider an ensemble
of disordered metallic grains in  
which electrons move in the presence of a random impurity potential
$V({\bf r})$ which is taken to be white noise random potential
with the variance $\langle V({\bf r})V({\bf r'})\rangle= \delta({\bf
  r}-{\bf r'})/2\pi \nu \tau$, where $\nu$ is the density of states.
The correlator of the thermodynamic potentials at different
values of the gate voltage $q$ and $q'$ can be written as 
\begin{equation}
  \label{eq:cumulants}
  \langle \Omega(q) \Omega(q')\rangle=T^2
\frac{\partial^2}{\partial \alpha\partial \alpha'}
      \langle  Z^\alpha(q) Z^{\alpha'}(q') \rangle \left.
\right|_{\alpha,\alpha'\rightarrow 0}, 
\end{equation}
Here $\langle Z^\alpha(q) Z^{\alpha'}(q') \rangle$ is the ensemble
averaged product of the replicated partition functions at different
values of the gate voltage.  For $\delta \ll T \ll E_C$ it can 
be expressed as a functional integral over the
$\hat{Q}$-matrices, arising from disorder averaging, and over the
phases $\phi_j$ which arise from decoupling the Coulomb
interaction (\ref{eq:cint}) via the
Hubbard-Stratonovich transformation, as follows:
\begin{mathletters}
\label{maineq}
\begin{eqnarray}
&&\langle Z^\alpha(q)Z^{\alpha'}(q')\rangle =\sum\limits_{\{W\}} \langle 
Z^{\alpha}Z^{\alpha'}\rangle_{_{W}} \, e^{-2\pi i
  \sum\limits_{j=1}^{\alpha+\alpha'}  q_j W_j}, 
\label{winding} \\
&&\langle 
Z^{\alpha}Z^{\alpha'}\rangle_{_{W}} =\int d\hat{Q}d[\phi]
e^{ -F[\hat{Q},\phi ] 
-\sum\limits_{j=1}^{\alpha+\alpha'}\left[\frac{ E_C q_j^2}{T} +
\int\limits_0^{\beta}\frac{\dot{\phi}_j^2 d\tau}{4E_C}\right]},
\label{integration} \\
&&F [\hat Q, \phi] = {\rm Tr} \left [ \frac{\pi \nu}{4\tau} \hat Q^2 -
  \ln \left( i\hat{J} +\frac{\vec{\partial}^2}{2m}+\frac{i\hat{Q}}{2\tau}   
  \right)\right]  .
\label{eq:freeen}
\end{eqnarray}
\end{mathletters}
Equations (\ref{maineq}) were obtained by taking the integral over the
static component of the field $\dot{\phi}$ by the saddle point
approximation which is valid for $T \gg \delta$~\cite{Gefen}. 
The resulting expression, Eq.~(\ref{winding}), is written as a sum over
the sets of  winding numbers $\{W\}$\cite{AES} which 
label  distinct topological classes of the phase,   
$\phi_j(\beta) =\phi_j(0)+2\pi W_j$.  
The treatment of the cooperon degrees of freedom requires doubling the
number of fermion fields. The covariant derivative $ \vec{\partial}$
in Eq.~(\ref{eq:freeen}) is expressed through the time-reversal
symmetry breaking matrix $\tau_3$~\cite{Efetov97} as $ \vec{\partial}
= \nabla +i e {\bf A}\hat{\tau}_3/c$, where ${\bf A}$ is the vector
potential, and ${\rm Tr}$ denotes the trace taken over the spin,
$\sigma$, coordinate, $r$, replica, $i,j$, and Matsubara, $n,m$, as
well as the time-reversal indices.  The matrix $\hat{J}$ in
Eq.~(\ref{eq:freeen}) is defined as 
\begin{equation}
\hat{J}\equiv \left[\hat{\varepsilon} -\frac{i q_j \delta}{2} +\Gamma
  e^{-i\phi_j(\tau)}\Lambda _{\tau,\tau'} e^{i\phi_j(\tau')}
  \right]\delta_{ij} 
  \delta_{\sigma, \sigma'}, 
\label{J}
\end{equation}
where $\hat \varepsilon = i \partial_\tau$ is diagonal in the
Matsubara frequencies, $\hat \varepsilon = \delta_{nm} \varepsilon_n=
\delta_{nm} \pi T(2n+1) $, and  $\Lambda
_{\tau,\tau'} =-i/\sin[\pi T(\tau - \tau')]$. 
 Finally, $q_j=q$ for $1\leq j \leq \alpha$ and
$q_j=q'$ for $\alpha < j \leq \alpha +\alpha'$.

At frequencies smaller than the inverse mean free time $1/\tau$ the
effective low-energy theory may be written as a functional integral
over the saddle point manifold of
$\hat{Q}$-matrices~\cite{Finkelstein83}. The low-energy degrees of
freedom are usually parameterized as $\hat{Q}=U\hat{Q}_0 U^{-1}$, where
the matrix $\hat{Q}_0$ is diagonal in the Matsubara space, and
represents the saddle point of the free energy $F[\hat{Q},\phi ]$
in Eq.~(\ref{eq:freeen}) at $\hat \phi =0$.  At frequencies below the
Fermi energy it is
given by $\hat{Q}_0=\Lambda$.  For our
present purposes, however, it is convenient to write the $\hat
Q$-matrix as $\hat{Q}=\hat{T}\hat{\lambda} \hat{T}^{-1}$, where
$\hat{\lambda}$ is 
the saddle point with the lowest free energy $F[\hat{Q},\phi ]$ for
a given $\hat \phi$~\cite{Kamenev98}.  It is given by the equation
\begin{equation}
   \label{eq:sp}
   \hat{\lambda} =\frac{i}{\pi \nu}\int \frac{d^dp}{(2\pi)^d} \left(
     i\hat{J} +\frac{(p- \frac{e{\bf A}\hat{\tau}_3}{c}) 
  ^2}{2m}+\hat{\mu}(\hat{q}) +\frac{i\hat{\lambda}}{2\tau}    
\right)^{-1}. 
\end{equation}
Note that since the operator $\hat{J}$ is coordinate-independent, the
saddle point matrix $\hat{\lambda}$ is uniform in space. This is a
consequence of the constant interaction model, Eq.~(\ref{eq:cint}). 

At $\phi_j(\tau)=2\pi T\tau W_j$ 
the matrix $\hat J$ is diagonal in the basis
of Matsubara frequencies. Since it is Hermitian it may be diagonalized
in some different basis for other configurations of
$\phi_j(\tau)$ as well: $\hat J=\delta_{nm} \delta_{ij}\delta_{\sigma,
  \sigma'}J^j_n$, 
where $n,m$ can be identified with the Matsubara indices only at
$\phi_j(\tau)=2\pi T\tau W_j$.  The saddle  point  
 $\hat{\lambda}$, Eq.~(\ref{eq:sp}) is also diagonal in this basis 
and  in the low-frequency sector is given by $\hat{\lambda} = {\rm
  sgn}\varepsilon_n \delta_{nm} \delta_{ij}\delta_{\sigma, \sigma'}$,
where $\varepsilon_n$ should be understood as
the Matsubara frequency corresponding to $J^j_n$ at $\phi_j(\tau)=2\pi
T\tau W_j$.

At $T \gg \delta$ the integration over the soft modes of $\hat{Q}$ in
Eq.~(\ref{integration})  may be carried out by expressing the rotation
matrices $\hat{T}$ in terms of the 
generators, $\hat{T}=\exp(iP)$, and expanding the free energy
$F[\hat{Q},\phi]$ to second order in $P$:
\begin{eqnarray}
  \label{eq:expansion}
F[\hat{Q},\phi]=&& F[\hat{\lambda},\phi] +\frac{\pi}{2 \delta} 
\sum \limits_{\rho=0,1}
\sum \limits_{\sigma,\sigma'}
\sum \limits_{i,j;n,m}'  |P_{ki;\sigma\sigma'}^{mn}(\rho)|^2
\nonumber \\ 
&& \times
\left(D{\bf p}^2 + J^i_n-
J^j_m +\rho/\tau_H\right). 
\end{eqnarray}
The prime over the sum in Eq.~(\ref{eq:expansion}) and below indicates
that the summation goes only over such pairs $n, m$ that
$\varepsilon_n >0$ and $\varepsilon_m <0$.  The Gaussian integration
over $P$ in Eq.~(\ref{integration}) then results in the following
expression
\begin{equation}
\Delta_{_{\{W\}}}=\prod\limits_{i,j;\rho}
\prod\limits_{{\bf p},m,n}'
\frac{\delta^4}{ (J^i_n-J^j_m 
+D{\bf p}^2+\rho/\tau_H)^4},
\label{diffuson}
\end{equation}
where the power $4$ arises from taking the product over the spin
indices.

The largest term in the sum over $\{W\}$ in Eq.~(\ref{winding}) has
a monotonic dependence on the gate voltages, $q$ and $q'$, and
corresponds to $\{W\}=\{0\}$ with all $W_j=0$. The leading oscillatory
contributions to this sum arise from the terms with $\{W\}=\{l,l'\}$,
with $l,l'=\pm 1$, having only one non-zero $W_j=l$ for $1\leq j \leq
\alpha$ and only one non-zero $W_j=l'$ for $\alpha < j \leq
\alpha+\alpha'$.  The dissipative free energy $F[\hat{\lambda},\phi]$ in 
Eq.~(\ref{eq:expansion}) has a minimum when $\phi_j=0$ 
in the replicas with $W_j=0$. In the replica with $W_j=l$ 
the instanton phase configuration~\cite{Korshunov} which minimizes  
the free energy may be written in terms of a complex variable
$z$~\cite{Nazarov}, 
\begin{equation}
  \label{eq:inst}
 \exp(i\phi_z)= 
\frac{u-z}{1-uz^*}, \qquad u=\exp(2\pi i T\tau ),
\end{equation}
whereas for $W_j=l'$  $\phi_j$ is given by Eq.~(\ref{eq:inst}) with 
$z \to z'$. The free energy $F[\hat{\lambda},\phi] = g_T$ on
the instanton configuration (\ref{eq:inst})  and is 
independent of $z$ and $z'$  which 
parameterize the widths and the positions of the 
instantons in replicas with $W_j=l,l'$ and satisfy the inequalities: 
$|z|,|z'|<1$ for $l,l' = 1$, and $|z|,|z'|>1$ for $l,l'= - 1$. 
We need to integrate over the fluctuations of $\phi$
around the instanton configuration Eq.~(\ref{eq:inst}). To this end we
write $\phi=\phi_z + \tilde \phi$, where $\tilde \phi$ represents the
massive modes. The integration over the zero modes should be
performed with the measure $d^2z/(1-|z|^2)$ for $|z|<1$. For $|z| >1$ 
the inversion transformation $z \to 1/z$ again maps the integration domain 
to $|z|<1$ with the integration measure $d^2z/(1-|z|^2)$.
All other fluctuations $\tilde \phi$ have a large mass of order $g_T$ and
can be integrated out  in the gaussian approximation.  Moreover, in
carrying out this integration one may replace the determinants
$\Delta[\phi]$ in Eq.~(\ref{diffuson}) by their values $\Delta[\phi_z]$ on the
instanton configurations (\ref{eq:inst}) due to their weak dependence on
$\tilde \phi$.  We then obtain
\begin{eqnarray}
&& \frac{\langle Z^{\alpha}(q)  Z^{\alpha'}(q') \rangle}{\langle
  Z^{\alpha}(q)  Z^{\alpha'}(q') \rangle_{\{0\}} }=
1 +  \alpha\alpha'  \sum \limits_{l,l' =\pm 1} e^{2\pi i(lq+l'q')}
\nonumber \\ && \times
  \int\limits_{|z|,|z'|<1}\frac{d^2zd^2z' f(z)f(z') }{(1-|z|^2)(1-|z'|^2)} 
  \frac{\Delta_{\{l,l'\}}[\phi_z]}{\Delta_{\{0\}}[0]},
\label{eq:Zz}
\end{eqnarray}
where $f(z)$ is given by the following expression~\cite{Beloborodov}
\begin{equation}
  \label{eq:fresult}
f(z)=\frac{g_T^2E_C \exp\left\{-\frac{2\pi^2
        T|z|^2}{E_C(1-|z|^2)}\right\}}{2\pi^3 T\left[1+\case{\Gamma}{\pi
T}\left( 1-|z|^2\right )\right]}. 
\end{equation}
With the aid of Eq.~(\ref{diffuson}) the determinant
$\Delta_{\{l,l'\}}[\phi_z]$  in Eq.~(\ref{eq:Zz}) can be 
expressed through the eigenvalues of the operator $\hat{J}$ on the instanton
configuration (\ref{eq:inst}). 
To this end it is convenient to express the operator
$\hat{J}$, Eq.~(\ref{J}), in terms of the complex variable $u$ in
Eq.~(\ref{eq:inst}), 
\begin{equation}
\label{Jcomplex}
\frac{\hat{J} +\frac{i q\delta}{2}}{2\pi T}=-u\partial_u
-\frac{i\Gamma}{2\pi^2T} 
\oint\frac{du'}{u'}\frac{1-uz^*}{u-z}\frac{\sqrt{uu'}}{u-u'}
\frac{u'-z}{1-u'z^*}.
\end{equation}
The integral in this equation is taken over the unit circle, $|u'|=1$,
and the operator $\hat{J}$ is  acting in the space of functions
spanned by the fermionic Matsubara frequencies $u^{-n-1/2}$.  
In the case of the instantons, $l,l'=1$, (anti-instantons, $l,l'=-1$)
the functions $u^{-n-1/2}$ with negative (positive) Matsubara frequencies,
are eigenfunctions of $\hat{J}$ 
with eigenvalues $J_n  =\varepsilon_n-\Gamma - iq\delta/2$. 
All other eigenvalues $J_n$ can be found from the following
equation~\cite{Beloborodov}
\begin{equation}
\label{sum}
\frac{1}{1-|z|^2}=\sum\limits_{k=0}^{\infty}
\frac{2\Gamma |z|^{2k}}{\mp (J_n +i q \delta /2 ) +\varepsilon_k + \Gamma },
\end{equation}
where the top (bottom) sign in the right hand side describes the 
instanton (anti-instanton) case.  For $\delta |q-q'| \gg \Gamma +T$ the
integral in Eq.~(\ref{eq:Zz}) is dominated by the region of short
instantons, $1-|z|^2 \ll \case{T}{T+\Gamma}$ and $1-|z'|^2 \ll
\case{T}{T+\Gamma}$.  
In this case the solutions of Eq.~(\ref{sum}) are given by
$J_n  = -i \delta q /2 \pm [\varepsilon_n+\Gamma  -2\Gamma
|z|^{2n}(1-|z|^2)]$, $n \geq 0$.   
Next, we use these eigenvalues to evaluate the determinant in
Eq.~(\ref{diffuson}).  Since the second term in
Eq.~(\ref{eq:Zz}) is multiplied by 
$\alpha \alpha'$, to obtain the correlator in Eq.~(\ref{eq:cumulants})
we only need to evaluate the ratio of the determinants in the second term
in the limit $\alpha, \alpha' \to 0$. It is not difficult to show,
see appendix C of Ref.~\cite{Beloborodov}, that in the limit $\alpha,
\alpha' \to 0$, $\Delta_{\{0\}}[\phi_z]$, $\Delta_{\{1,1\}}[\phi_z]$ and
$\Delta_{\{-1,-1\}}[\phi]$ are equal to unity, and 
$\Delta_{\{1,-1\}}[\phi_z]$ is given by
\begin{equation}
\label{det2}
\frac{\Delta_{\{1,-1\}}}{\Delta_{\{ 0\}}}=e^{ \; \; \sum\limits_{
\rho,{\bf p}} \; \sum\limits_{ n,m\geq 0 } 
\frac{16\Gamma^2 |z|^{2n}(1-|z|^2) |z'|^{2m}(1-|z'|^{2})}{(\varepsilon_n+ 
\varepsilon_m + D{\bf p}^2 + a_\rho )^2}},
\end{equation}
with $a_\rho$ defined below Eq.~(\ref{irreducible}). The ratio 
$\case{\Delta_{-1,1}[\phi_z]}{\Delta_{\{ 0\}}}$ is given by the complex
conjugate of  
Eq.~(\ref{det2}).  Using  Eqs.~(\ref{det2}), (\ref{eq:Zz}) 
and (\ref{eq:cumulants})  we obtain Eq.~(\ref{2}).

In conclusion we note that: i) Equation (\ref{2}) was obtained with
logarithmic accuracy, and is valid for
temperatures $\delta \ll T\ll E_C$. 
To derive it we also neglected the instanton-instanton
interactions, which is valid for $T \gg g_T^2 E_C
\ln\frac{E_C}{\Gamma}\exp(-g_T/2)$. 

ii) The results for the orthogonal and the unitary ensembles are
obtained from Eq.~(\ref{2})  by taking the
limits $\tau_H \to \infty$ and $\tau_H \to 0$ respectively. If the
Zeeman splitting is taken into account the degeneracy of
diffusons/cooperons in Eq.~(\ref{det3}) corresponding to different
spin projections of the particles is lifted: two of them remain
unchanged, and the other two acquire a correction $i \delta ( q -q')
\to i \delta ( q -q') \pm i 2 \mu H$, where $\mu$ is the Bohr
magneton.

iii) The correlator of differential capacitances, or thermodynamic
densities of states, may be obtained from Eq.~(\ref{irreducible}) by
differentiating it with respect to $q$ and $q'$.  The correlations fall
off as $|q-q'|^{\frac{d}{2}-2} \ln ^2 |q-q'|$  at $|q-q'|\gg E_T/\delta$.
Equations  (\ref{irreducible}) and (\ref{det3}) demonstrate the inadequacy
of the mean-field 
treatment of the Coulomb interaction for the description of the
Coulomb blockade oscillations.  Under such treatment the
Coulomb interaction in the grain would merely result
in the linear $q$-dependence of the self-consistent chemical potential
for the non-interacting problem.  In contrast with Eq.~(\ref{2}) 
the corresponding oscillatory part
of the correlator of densities of states would decay
exponentially  $\propto |q-q'|^{-4} \exp \left\{- \gamma_d \left( \case{
\delta |q-q'|}{ E_T}\right)^{d/2}\right\}$~\cite{Andreev95}, where
$\gamma_d$ is a numerical coefficient.

iv) The mesoscopic fluctuations of the Coulomb blockade
oscillations in the strong tunneling regime were first studied in
Ref.~\cite{Aleiner97}. The single particle
properties of the dot in this approach are described in terms of the
scattering matrix at the contact. The long range correlations obtained
in Ref.~\cite{Aleiner97} resulted from neglecting the dependence of 
the scattering matrix on the gate voltage.

v) In the $\sigma$-model approach the Coulomb blockade oscillations are
described by the instantons of the $\phi$-field~\cite{Korshunov}.
Their mesoscopic fluctuations  are
described  by the fluctuations of the $Q$-matrix about the instanton
configurations. The problem therefore reduces the evaluation of
diffusons on the instanton configurations of the $\phi$-field.

We thank I.~Aleiner, A.~Kamenev, A.~Larkin, K.~Matveev and 
Yu.~Nazarov for helpful discussions. 
We gratefully acknowledge of the Ruhr-Universit\"{a}t,
Bochum, Germany, where part of this work was performed.
This research was sponsored by the Grants DMR-9984002,
BSF-9800338 and SFB 237, and by the A.P. 
Sloan and the Packard Foundations.

\end{multicols}
 
\end{document}